\documentclass[prb,preprint,showpacs]{revtex4}
\usepackage{graphicx}
\usepackage{color}

\begin{document}

%opening
\title{Doping of C$_{60}$-induced electronic states in BN nanopeapods}

\author{Vladimir Timoshevskii}
\affiliation{Department of Physics, McGill University, 3600 rue University, Montr\'eal, Qu\'ebec, Canada H3A 2T8}
\email{vladimir@physics.mcgill.ca}

\author{Michel C\^ot\'e}
\affiliation{D\'epartement de physique, Universit\'e de Montr\'eal et Regroupement qu\'eb\'ecois sur les mat\'eriaux de pointe (RQMP) Universit\'e de Montr\'eal, Case Postale 6128, Succursale Centre-ville, Montr\'eal, Qu\'ebec, H3C 3J7 Canada}
\email{Michel.Cote@umontreal.ca}

\begin{abstract}
We report the results of \textit{ab initio} simulations of the electronic properties of a chain of C$_{60}$ molecules encapsulated in a boron nitride nanotube - so called BN-nanopeapod. It is demonstrated that this structure can be effectively doped by depositing potassium atoms on the external wall of the BN-nanotube. The resulting material becomes a true metallic one-dimensional crystal, where the conduction states are formed solely by the fullerene chain. At the doping rate of one K atom per C$_{60}$ molecule, the system shows the density of states at the Fermi level considerably higher than in any of the fullerene crystals presently made. This makes the doped BN-peapod structure an interesting candidate to study a possible superconducting state.
\end{abstract}

\pacs{73.22.-f 71.15.Mb 74.70.Wz}

\maketitle

\section{Introduction}

Since it was first synthesized by Smith {\it et al.} \cite{Smith98} in 1998, a chain of fullerene molecules, encapsulated in a single wall carbon nanotube, - so called ''carbon peapod'' - has attracted considerable attention of both experimentalists and theoreticians. This combination of two carbon allotropes represents a new class of hybrid materials, showing interesting physical properties. The encapsulated fullerene chain influences significantly the local electronic structure of the host nanotube, changing the value of the band gap, as well as the density of electronic states. As a result, physical properties such as electric resistance and thermopower are different for the peapods and for the host carbon nanotubes \cite{Hirahara00,Vavro02}. A photoinduced electron transfer effect has been recently observed in C$_{60}$ nanopeapod devices, making them promising candidates for nanoscale photodetectors \cite{Li08}. The peapods are also promising materials for a possible application in the area of high-temperature superconductors. Being encapsulated in the nanotube, the fullerene molecules form a one-dimensional crystal, which is expected to show an enhanced density of conduction electronic states, as compared to three-dimensional fullerene systems \cite{Service01}.

The use of a boron-nitrogen (BN) nanotube as a host for a fullerene chain was first theoretically proposed by Okada \textit{et al} \cite{Okada01}. The synthesis of this nanomaterial was soon after realized by Mickelson \textit{et al} \cite{Mickelson03}. Their electronic and structural properties were calculated in a few theoretical studies \cite{Kang04, Trave04, Hill07}. Due to its unusual electronic structure, the C$_{60}$@BN is an interesting candidate to study a possible superconducting state. Compared to carbon nanotubes, which can be metallic or semiconducting depending on diameter and chirality, the BN-nanotubes are always insulating with an energy gap of about 4 to 5 eV, independent of the nanotube geometrical configuration. Therefore the lowest conduction bands of the boron-nitrogen nanopeapod should essentially be formed by the LUMO states of the C$_{60}$ which emerge within the energy gap of the BN nanotube \cite{Okada01}. After being doped, this system is expected to conduct via the fullerene t$_{1u}$-originating bands, as in the case of the doped three-dimensional fullerene crystals \cite{Gunnarsson97}. However, to our knowledge, no effective method of doping of the BN-peapod has been proposed yet, neither experimentally nor theoretically. Finding such a method is crucial for future experimental studies of the electronic and superconducting properties of this nanomaterial.

In this study, we propose a simple and effective method of doping the BN-peapod structure. Our theoretical results, based on \textit{ab initio} calculations, show that the BN-peapod can be doped externally, without the introduction of alkali atoms inside the BN nanotube. The proposed method has two main advantages. First, it does not destroy the internal order of the BN-peapod, which is crucial to favor the superconducting state in this system, and second, it should be rather simple for an experimental implementation.

\section{Calculation details}

Our calculations were performed within the local density approximation (LDA) \cite{CA-PZ} of the density-functional theory (DFT) \cite{HK-KS}. The valence electrons were treated explicitly, while the interactions of the core electrons with atomic nuclei were described by norm-conserving pseudopotentials \cite{TM-KB}. We used the SIESTA program package \cite{SIESTA}, which is a self-consistent DFT code employing numerical atomic orbitals (NAO) as a basis set. An effective basis set, consisting of doubled \{$s,p_x,p_y,p_z$\} orbitals plus polarization orbitals of $d-$type was used. A shifted 16 {\bf k}-point grid was used to sample the Brillouin zone (BZ) for the empty nanotube calculations whereas a shifted 4 {\bf k}-point grid showed good results for structural relaxation of the peapod. A supercell approach was adopted in the directions perpendicular to the tube axis. The wall to wall distance between the nanotube and its images in the neighboring unit cells was fixed at 16 \AA~to minimize possible interactions. The self-consistency cycle was considered converged when the total energy difference between two iterations was less than 1 meV, and the geometry relaxation was stopped when the maximum atomic force in the system was less than 0.02 eV/\AA. The calculations were performed in a non-spin-polarized approximation.

\begin{figure}
  \begin{center}
  \includegraphics*[scale=0.35]{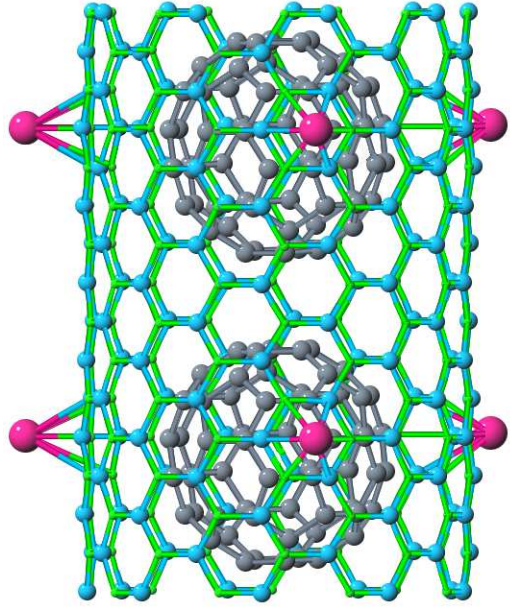}
  \caption{\label{peapod} (Color online) The geometry of the C$_{60}$@K$_{3}$BN(9,9) peapod. The large circles show the K atoms, absorbed on the nanotube wall. Figure shows a doubled unit cell along the tube axis.}
  \end{center}
\end{figure}

We used a (9,9) armchair BN-nanotube to construct a peapod. First, the empty nanotube was fully relaxed with respect to both the atomic positions and the cell size along the nanotube axis. The obtained lattice parameter of 2.49 \AA~is in agreement with previous theoretical calculations \cite{Okada01,Guo05}. Afterward, a peapod was constructed applying commensurability conditions between the periodicity of the nanotube and that of the chain of the fullerene molecules. The unit cell of the resulting peapod structure, see Fig.\ref{peapod}, consists of one C$_{60}$ molecule and a part of a nanotube, which corresponds to a quadruple period of a BN-nanotube. The atomic positions were relaxed while keeping the lattice parameter along the nanotube axis fixed. The resulting structure has a fullerene every 9.96 \AA. The same interfullerene distance (9.96 \AA) was previously obtained by relaxing a standalone weakly interacting fullerene chain \cite{Trave04}, which shows that the chosen commensurability condition is well suited to describe the present structure.

As an additional quality test of the NAO basis set of potassium, boron and nitrogen atoms, we performed a series of test calculations of potassium atom, deposited on a single boron-nitrogen monolayer. The system was completely relaxed, and its electronic structure was analysed. The same calculation was performed using a highly accurate Linearized Augmented Plane Wave (LAPW) method \cite{wien2k}. This method is of all-electron type, and does not use any pseudopotentials. The pseudopotential calculations showed that the K atom relaxes at 3.52~\AA~ from the BN surface, which is in good agreement with all-electron result of 3.57~\AA. Also, both methods showed a zero charge transfer from the potassium atom to the boron-nitrogen monolayer. These results ensure that the qualities of pseudopotentials and the NAO basis set are sufficient to use them in the present study of the more complex BN-peapod structure.

\section{Results and discussion}

\begin{figure}
  \begin{center}
  \includegraphics*[width=11cm]{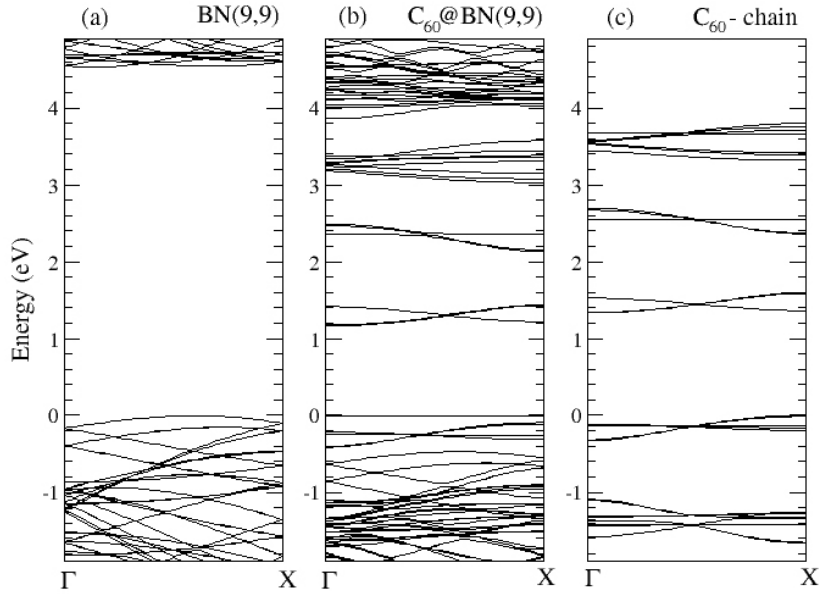}
  \caption{\label{bands} The calculated band structure of (a) BN(9,9) nanotube, (b) C$_{60}$@BN(9,9) peapod, (c) free-standing linear chain of C$_{60}$ molecules. The zero of energy scale corresponds to the valence band maximum.}
  \end{center}
\end{figure}

The calculated band structure of the C$_{60}$@BN peapod is presented in Fig.\ref{bands}(b) which is in agreement with the previous theoretical study \cite{Okada01}. Comparing the band structure of the peapod with the empty BN(9,9) nanotube, Fig.\ref{bands}(a), we see that the fullerene-induced bands emerge into the band gap of the nanotube. The combined structure becomes a semiconductor with a band gap of about 1.2 eV (DFT-LDA value). We also observe that both the valence band maximum (VBM) and conduction band minimum (CBM) states of the peapod are formed solely by the C$_{60}$-induced bands, originating from the fullerene $h_u$- and $t_{1u}$-states respectively. To study the influence of the nanotube on the electronic structure of the encapsulated fullerene chain we present in Fig.\ref{bands}(c) the calculated band structure of the chain of the C$_{60}$ molecules without the BN nanotube host. The distance between the fullerene molecules was kept the same as in the peapod, while the atomic positions were allowed to relax. We observe that the effect of the encapsulation is practically negligible for the first and second group of the fullerene conduction bands, and the band gap is reduced by 0.1 eV in the BN-peapod as compared to the isolated C$_{60}$ chain due to small changes in the topology of the $h_u$ derived valence band maximum in the peapod. These results allow to conclude that, contrary to the carbon nanotube peapods, the BN nanotube introduces minimum changes in the electronic structure of the fullerene-induced bands. Hence, the BN-peapod can be considered as a real one-dimensional fullerene crystal, where the C$_{60}$ molecules are constrained to form a linear chain by means of a weak van der Waals interaction with the BN nanotube host.

Finding a simple and effective method of doping of the BN-peapod structure is crucial for further progress in studying the electronic and superconducting properties of this system. By analogy with intercalated carbon peapods \cite{Guan05}, the BN-peapods can probably be doped by encapsulating potassium atoms inside the nanotube. However, this method influences significantly the positions of the intercalated C$_{60}$ because of the unavoidable random distribution of potassium atoms inside the nanotube. Moreover, the interaction of the intercalated potassium atoms with a host nanotube was shown to be substantially different for the cases of carbon and BN nanotubes. If in the case of carbon nanotubes, the potassium valence electrons are donated to the nanotube wall, for BN nanotubes, the potassium {\it s-} state remains occupied and the energetically preferable potassium atom position is at the center of the tube \cite{Rubio96}.

In the present study we investigated, with the help of electronic structure simulations, a simple possible approach to dope a BN-peapod : external deposition of potassium atoms on the wall of the BN-nanotube. Although this method is free from the disadvantages of the internal encapsulation approach, another problem may appear due to a large separation between the fullerene chain and the potassium atoms. In this case, the potassium {\it s-}electrons should not only be donated to the wall of the nanotube, but should be transferred further to occupy the conduction states localized on the C$_{60}$ chain. We studied the energetics and the electronic structure of the BN-peapod with one, two and three potassium atoms, deposited on the external wall of the nanotube, as shown in Fig.\ref{peapod}. The doping rate corresponds to one, two and three electrons per C$_{60}$ molecule respectively. For each case, the structure was fully relaxed, and the results indicate that the potassium atoms attach themselves to the wall of the nanotube. They occupy an equilibrium position on top of the BN-hexagon with a K-B(N) distance of 3 \AA. To confirm that the potassium atom energetically prefers to be deposited on the nanotube wall, we calculated the heat of formation of this structure. The heat of formation was calculated by subtracting the total energy of the doped peapod from the sum of the total energies of separated systems of bulk potassium and the undoped peapod. The results show that the external deposition of the K atoms on the BN-peapod wall is an exothermic reaction with the heat of formation of 0.56 eV/K atom. This value is close to 0.5 eV/K atom, obtained in a previous study for deposition of a K monolayer on a planar BN sheet \cite{Rubio96}.

\begin{figure}
  \begin{center}
  \includegraphics*[width=11cm]{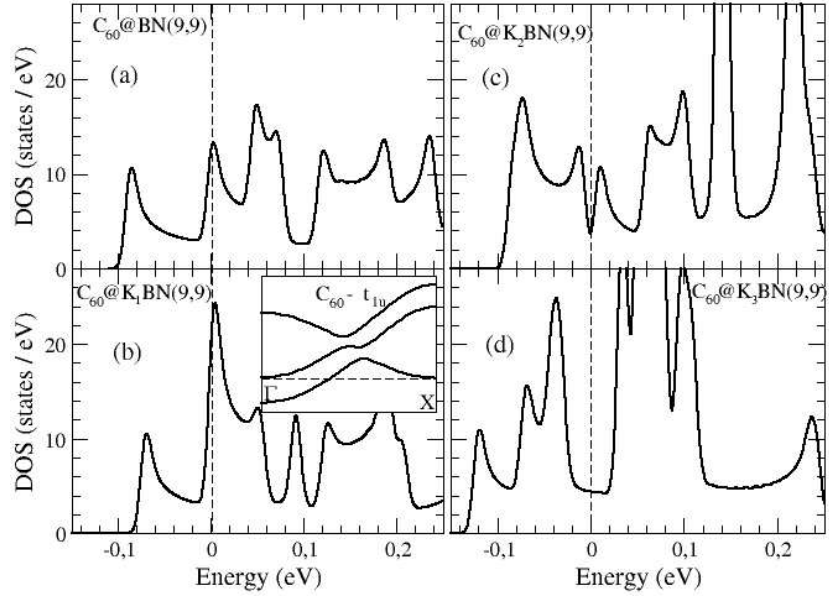}
  \caption{\label{dos} Calculated density of electronic states for (a) the C$_{60}$@BN(9,9) peapod, doped by one electron within a ''rigid band'' model approach, (b,c,d) the C$_{60}$@K$_{x}$BN(9,9) peapod, doped by one, two and three potassium atoms respectively. The dashed line corresponds to the Fermi level position. The inset in (b) shows the topology of electronic bands near the Fermi level for C$_{60}$@K$_{1}$BN(9,9) peapod.}
  \end{center}
\end{figure}

\begin{figure}
  \begin{center}
  \includegraphics*[angle=-90,width=12cm]{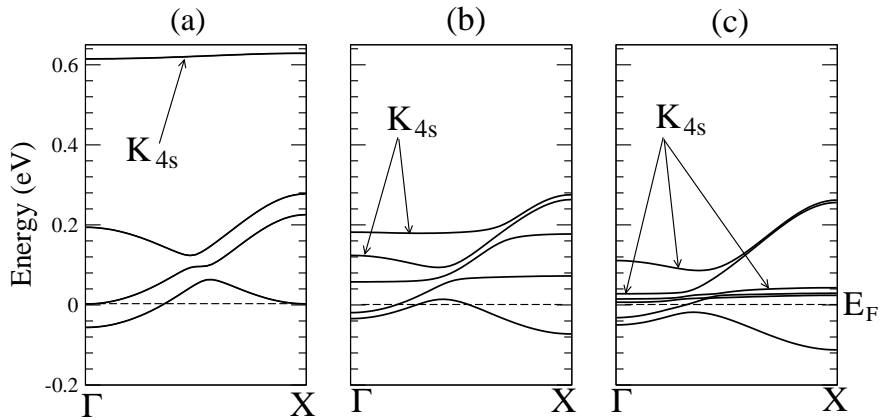}
  \caption{\label{bandsEF} The band structure of C$_{60}$@BN(9,9) peapod under different potassium doping. The panels (a), (b), and (c) correspond to the doping rate of 1, 2, and 3 K-atoms per C$_{60}$ molecule respectively. The dashed line corresponds to the Fermi level position.}
  \end{center}
\end{figure}

We started our study of the effect of potassium doping on the electronic structure of the BN-peapod by calculating the density of electronic states (DOS) of the doped structure. We compare in Fig.\ref{dos}(b,c,d) the DOS of the first group of conduction bands of the BN-peapod, doped by one, two, and three potassium atoms per C$_{60}$ molecule respectively. The corresponding DOS for the uncharged C$_{60}$@BN(9,9) peapod where the geometry of the system was fixed to the relaxed C$_{60}$@K$_1$BN(9,9) is also presented in Fig.\ref{dos}(a). In this last case, the Fermi level is set within a ''rigid band'' model approach to one electron per fullerene. This allows us to see the pure effects of the charge transfer on the electronic structure of the system. The results show that under the K- doping the BN-peapod becomes metallic, and the potassium 4$s$- electrons occupy the t$_{1u}$-derived fullerene bands. However, calculations demonstrate that the application of a ''rigid band'' model is questionable in the case of a BN-peapod structure. By comparing Fig.\ref{dos}(a) and (b) we see that the transfer of only one electron to the conduction states leads to significant changes in the density of electronic states: the first peak in the DOS practically does not change, while the amplitude of the second one is almost double in the case of potassium doping. The DOS profile of the conduction states, located above the Fermi level, is also substantially different in case of a ''rigid band'' and potassium doping. For a doping one K atom per C$_{60}$, the conduction states of the peapod are still formed by t$_{1u}$ fullerene induced bands (the inset in Fig.\ref{dos}(b)), while if the doping is further increased, the K-induced states shift down in energy and start to hybridize with the states of the C$_{60}$ molecule, changing dramatically the density of conduction electronic states of the peapod (Fig.\ref{dos} (c,d)). The effect of the influence of the K-induced states on the t$_{1u}$-derived fullerene bands with increase of the doping concentration is explicitly demonstrated in Fig.\ref{bandsEF}. We see that at the doping rate of one K-atom per C$_{60}$ molecule, the empty potassium 4$s$ state lies at 0.6 eV above the Fermi level (Fig.\ref{bandsEF}(a)). With a increase of the doping level, the potassium states are pushed down in energy, and start to mix with the fullerene states (Fig.\ref{bandsEF}(c,d)), causing the substantial changes in the DOS profile, shown in Fig.\ref{dos} (c,d). The present study also indicates that other dopants than potassium atoms will be necessary to effectively dope this system to concentrations higher than 1 electron per C$_{60}$ molecule that would preserve the fullerene band topology.

\begin{figure}[b]
  \begin{center}
  \includegraphics*[scale=0.6]{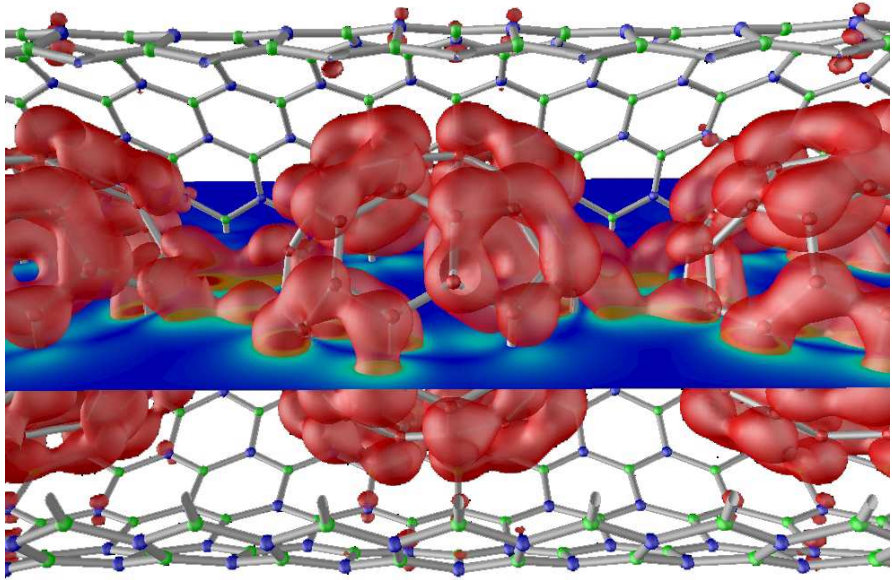}
  \caption{\label{charge} (Color online) Spacial distribution of the conducting electrons at the Fermi level for the C$_{60}$@K$_{1}$BN(9,9) peapod.}
  \end{center}
\end{figure}

Fig.\ref{dos}(b) demonstrates that being externally doped by one potassium atom, BN-peapod becomes a true metallic system with a high density of electronic states at the Fermi level. The conduction states at the Fermi level are formed only by the orbitals of the C$_{60}$ molecule, and the obtained DOS of 24 states/eV/spin \cite{spin} is considerably higher than 10 states/eV/spin, calculated for the FCC fullerene crystal within the same approach. This value is also considerably higher than a typical range of DOS (6.6-9.8 states/eV/spin), theoretically obtained for K$_3$C$_{60}$ crystal \cite{Gunnarsson97}. We should note here that the center to center interfullerene distance in the studied BN-peapod (9.96~\AA) is very close to the one in the FCC C$_{60}$ crystal (10~\AA), resulting in a band dispersion of the order of 0.5 eV for both materials. This demonstrates that a strong increase of DOS in the case of the BN-peapod is not caused by a reduction of the band dispersion, but is a consequence of the one-dimensional nature of the peapod structure, where the singularities at the Brillouin zone boundaries give rise to large peaks in the density of electronic states. This effect, which is absent in the case of three-dimensional systems, suggests that the BN-peapod structure can be more efficient in yielding a larger increase in DOS at the Fermi level than the doped fullerene crystals considered so far. To complete our study we calculated the spacial distribution of electronic density at the Fermi level for the C$_{60}$@K$_1$BN(9,9) system, which is presented in Fig.\ref{charge}. As expected from the DOS calculations, we observe that the conducting charge in the peapod, doped by one potassium atom, is localized on the C$_{60}$ subsystem. At the same time, the conducting electrons are delocalized within the fullerene chain, which ensures the possibility of electronic transport in this system.

To check the effects of possible spin polarization, we performed an additional calculation of the band structure of C$_{60}$@K$_1$BN(9,9) system using the local spin density approximation (LSDA). The initial spin polarization was artificially induced on all atoms. However, the final solution converged to a non-spin-polarized one, and the obtained band structure was completely identical to the one, presented in Fig.\ref{bandsEF}(a).

At the end of our study we briefly comment on the possibility of a superconducting state in C$_{60}$@BN system. A recent discovery of intrinsic superconductivity in narrow carbon nanotubes \cite{Tang01}, followed by a number of theoretical studies, has raised important questions, regarding superconducting properties of one-dimentional systems, and even questioned the validity of mean-field technique in describing this phenomenon \cite{Charlier07}. It is known that in 1D systems any macroscopic-scale phase order is suppressed by thermal and quantum fluctuations \cite{Mermin66}, which means that a pure superconducting state can be achieved only at $T=0 K$. However, in the temperature interval of $0<T<T_c$, the superconductivity is not completely destroyed, but appears in the form of superconducting fluctuations. This exact behavior was observed in Meissner effect measurements for narrow (4 \AA) single wall carbon nanotubes which exhibit a superconducting behavior at temperatures below 20K \cite{Tang01}. The smooth temperature variations of magnetic susceptibility in this system are due to its one-dimensional nature, and, by analogy, the same type of behavior may be expected from the peapod structure. Another well-known feature of 1D systems is the so-called charge-density-wave (Peierls) instability, which opens the gap on the Fermi level of a 1D metallic system due to the coupling of conduction electrons with low-energy phonons. This transition, which turns a metallic system into an insulating one, enters in direct competition with a superconducting transition, and once happened at higher temperatures than the latter, no superconductivity may be expected. However, for the doped BN peapods, the phonons responsible for the Peierls instability are intermolecular whereas it is the intramolecular modes than give rise to the coupling of the conducting electrons which yield the superconducting state. Since these modes have quite different energies, it might set apart the phase transition temperatures. This fact shows the importance of studying the low-energy excitation spectrum of C$_{60}$@BN systems and further investigations are necessary in this field to obtain a clear picture of the possibility of a superconducting state in peapod structures. 

\section{Conclusion}

We have presented a theoretical study of the electronic properties of the chain of the C$_{60}$ molecules, encapsulated in the boron-nitride nanotube - BN-peapod structure. We proposed a simple and effective method of doping the BN-peapod by depositing the potassium atoms on the external walls of the BN-nanotube. It has been demonstrated that under this type of doping the potassium valence electrons are transferred inside the peapod and occupy the conduction bands of the fullerene chain, turning the peapod into a metallic one-dimensional fullerene crystal. We have further demonstrated that being doped by one potassium atom per C$_{60}$ molecule, the system shows a density of electronic states at the Fermi level considerably higher than in any of the C$_{60}$ crystals presently made, which makes a BN-peapod structure a promising material to study superconducting properties in the fullerene-based materials.

% \bibliography{bn_peapod}

\begin{thebibliography}{21}
\bibitem[{spin()}]{spin}
\bibinfo{note}{As previously mentioned, the calculations were done in a non-spin-polarized
approximation. The values of DOS at the Fermi level are presented ``per spin`` to allow 
direct comparison with literature values for fullerene systems. A test spin-polarized 
calculation has also been performed (see later in text)}.
\end{thebibliography}

\newpage

\end{document}